# Reducing unnecessary alerts in pedestrian protection systems based on P2V communications

**Ignacio Soto** [1,*], **Felipe Jimenez** [2], **Maria Calderon** [1], **Jose E. Naranjo** [2] and **Jose J. Anaya** [2]

[1] Departamento de Ingeniería Telemática; Universidad Carlos III de Madrid; Av. Universidad, 30; 28911 Leganés (Madrid); Spain; isoto@it.uc3m.es (I.S.); maria@it.uc3m.es (M.C.)

[2] Instituto Universitario de Investigación del Automóvil (INSIA); Universidad Politécnica de Madrid; Campus Sur UPM; Carretera de Valencia, km. 7; 28031 Madrid (Madrid); Spain; felipe.jimenez@upm.es (F.J.); joseeugenio.naranjo@upm.es (J.E.N.), jj.anaya@upm.es (J.J.A.)

**\*** Correspondence: isoto@it.uc3m.es; Tel.: +34-916245974 (I.S.)

**Abstract:** There are different proposals in the literature on how to protect pedestrians using warning systems to alert drivers of their presence. They can be based on onboard perception systems or wireless communications. The evaluation of these systems has been focused on testing their ability to detect pedestrians. A problem that has received much less attention is the possibility of generating too many alerts in the warning systems. In this paper, we propose and analyze four different algorithms to take the decision on generating alerts in a warning system that is based on direct wireless communications between vehicles and pedestrians. With the algorithms, we explore different strategies to reduce unnecessary alerts. The feasibility of the implementation of the algorithms is evaluated with a deployment using real equipment, and tests are carried out to verify their behavior in real scenarios. The ability of each algorithm to reduce unnecessary alerts is evaluated with realistic simulations in an urban scenario, using a traffic simulator with vehicular and pedestrian flows. The results show the importance of tackling the problem of driver overload in warning systems, and that it is not straightforward to predict the load of alerts generated by an algorithm in a large-scale deployment, in which there are multiple interactions between vehicles and pedestrians.

**Keywords:** Vulnerable road users; pedestrian protection; traffic simulation; driver alert systems; pedestrian-to-vehicle (P2V) communications.

## 1. Introduction

One of the social needs related to road safety is the reduction in the number of accidents that involve vulnerable road users (VRUs). This group includes cyclists, motorcyclists and pedestrians, and their common feature is that, in an accident, they are not protected by a body vehicle. This means that the severity of their injuries is always higher than in the passengers of any other motor vehicle, which translates to a higher number of fatalities. In fact, the traffic accident statistics show that the number of road accidents involving VRUs is not decreasing at the same rate as other types of accidents [1]. These antecedents justify that one of the priorities of the research strategy of the European Union is the development of new information and communication technologies focused on improving the safety of VRUs.

Specifically, pedestrians are the most vulnerable group of VRUs, since they are the passive part of accidents and, in the event of an accident, their injuries are usually important. Using a mechanism to provide additional awareness to drivers seems to be a simple approach to improve the safety of pedestrians. Some systems with this aim have been proposed in the literature. These warning systems help drivers to detect, in a better way than using only their senses, situations in which there is a risk of an accident. The systems can be based on sensors or on a communication mechanism between vehicles and pedestrians. The evaluation of these warning systems usually focuses on measuring their detection ability in different scenarios.

A different problem that has received much less attention in the literature is the potential overload of drivers by a constant flow of alerts generated in the warning system. This problem is addressed in this paper, in which we explore strategies for a warning system to reduce unnecessary alerts. Unnecessary alerts are alerts generated by the warning system in situations that do not lead to a real danger for a pedestrian. In the design of a warning system, there is a trade-off between capturing all the risky situations and generating so many alerts that the system becomes useless because drivers cannot pay attention to the alerts.

In this paper, we consider the use of a cooperative warning system based on the connectivity of pedestrians and vehicles, using Person-to-Vehicle (P2V) short-range wireless communications. We propose four different alternative algorithms to activate alerts that warn drivers about potential risks to pedestrians. These algorithms represent different strategies to avoid unnecessary alerts, while still detecting all real risk situations for pedestrians. We validate the algorithms in real-life tests, involving real vehicles and pedestrians, to verify the feasibility of the implementation of the system. Then, to evaluate whether it is possible to reduce the alerts generated without losing alerts that represent a real danger for pedestrians, we carried out a comprehensive simulation study that allowed us to evaluate the behavior of the algorithms in an urban scenario.

The contributions of our paper are the following: (I) we identify the need to evaluate, in pedestrian protection systems, the unnecessary alerts generated by these systems; (II) we show that different algorithms to decide when to raise an alert in a warning system that behave well when evaluated in a small-scale situation, can behave very differently in terms of load on drivers when the system is used in vehicles roaming around urban areas; (III) we propose an algorithm to decide when to create alerts in a warning system to protect pedestrians that is implementable in real equipment and reduces unnecessary alerts by focusing on situations with high probability of risk to pedestrians; and (IV) we develop a simulation framework, based on the SUMO traffic simulator, which allows to accurately represent the interactions of vehicles and pedestrians in realistic urban scenarios. The simulation framework is an important tool to assess the effects of the deployment of warning systems in the real world.

The rest of the paper is organized as follows: section 2 reviews previous works on systems for VRU protection; section 3 presents the algorithms analyzed; section 4 describes the experiments with the real-life implementation of the algorithms; section 5 analyzes, by means of simulations, the ability of the different algorithms to reduce unnecessary alerts for drivers traveling through a city area; and finally, section 6 concludes the paper.

## 2. State of the art

The analysis of the pedestrian safety conditions [2,3] and the deployment of measures to solve this problem has been a common trend over the years. In the field of Advanced Driver Assistance Systems (ADAS), several systems have been proposed to reduce road accidents in the past. Most systems rely on onboard perception sensors such as computer vision and laser scanning (e.g. [4,5]). Both technologies have limitations and strengths [6,7]. Computer vision is especially capable of reconstructing a model of the environment, although problems arise in adverse weather and in changing lighting conditions. In addition, a high computational cost is required. On the other hand, laser scanners can be a powerful alternative with similar detection ranges. However, specific and complex algorithms are required, and the divergence of laser beams is a clear limitation for detecting pedestrians far away from the vehicle. These limitations motivate the use of sensor fusion in the perception of vehicle surroundings (e.g., [8,9]).

In the case of pedestrian detection, a special treatment is required due to the clear differences with other road users, such as their low speed, areas where they move, the ability to change direction, their small size compared with vehicles, their unprotected structure, etc. These characteristics imply that assistance systems should be developed specifically for this purpose. The first works focused on the use of computer vision, but even today, researchers are proposing novel approaches in this field (e.g., [10]) to overcome the limitations detected. With the deployment of laser scanners in vehicles for autonomous and semiautonomous driving functions, these sensors appear to be an alternative solution and satisfactory results have been found [11,12]. Finally, solutions based on fusion of sensors, computer vision and laser scanning, provide a promising area of research, even more so when these sensors are also applied in vehicles for many other purposes [13]. A recent research project, EU PROSPECT project [14], has followed this research line, investigating how to improve the effectiveness of active VRU safety systems based on video and radar technologies.

On the other hand, the current deployment of cooperative systems (C-ITS) provides vehicles with the possibility of accessing a large data set on what is happening in their short, medium and long range [15,16]. In this field, the use of Person-to-Vehicle (P2V) wireless communications has also been proposed to reduce the number of crashes involving VRUs, as a particular case of safety systems [17-29]. More specifically, the service called "Vulnerable road user protection (pedestrians and cyclists)" is included in the list of Day 1.5 C-ITS Services [30]. The EU VRUITS project [31] proposed an interesting architecture for the integration of VRUs in Cooperative ITS systems.

Nevertheless, the development of VRU protection systems based on wireless communications require to address several challenges, different from those being studied for systems based on on-board sensors, solving issues related to the correct pedestrian detection, the communications equipment involved, the network congestion caused by exchanged messages, the avoidance of false negatives or false positives, etc. A number of interesting research works have addressed these challenges. Some of them have focused on the possibility of using smartphones to establish Person-to-Vehicle communications, and transmit pedestrian's location data (i.e., beacons). The radio technologies considered are Wi-Fi direct [19], Wi-Fi [22], 802.11p [26,28] or cellular [24,25,27,29]. In addition, the new advances in 5G cellular networks, with the arrival of Device-to-Device operations [32] can open new communication alternatives that can be useful for pedestrian protection systems.

On the other hand, given the large number of pedestrians and vehicles in populated urban areas, a key issue that has been studied most recently is network congestion (e.g. [18,26]) caused when safety messages are exchanged by pedestrian and vehicles. These safety messages could be standard C-ITS messages (e.g., using ETSI ITS-G5 in Europe) [26,28]. These works show that the congestion of the wireless channel can have a significant impact on the loss and latency of messages. Therefore, there are different proposals to reduce network congestion, such as, to reduce the transmission power in pedestrians' smartphones [21] or to adapt the beaconing frequency, e.g., using a higher beaconing rate when a pedestrian is at a higher risk [18,21,26].

The study of pedestrian protection systems based on collision prediction algorithms has also received some attention from researchers [21,22,26,29]. The main issue with these algorithms is that their performance is very sensitive to pedestrians who change their course suddenly, a situation that can be common. In fact, more elaborate algorithms, using many parameters to predict whether a collision is going to happen or not, can exacerbate the problem and lead to false negatives. Additionally, the evaluation of these protection systems is of paramount importance. The most common approach to evaluate them has been to deploy small scenarios, for example, a single intersection; and the evaluation is limited to validate the functionality of the system, e.g., that alerts are generated according to the design of the solution.

A related problem that has received much less attention is the driver overload caused by the alerts generated by the protection system, and strategies to reduce unnecessary alerts. This issue is especially critical when drivers travel around cities with multiple pedestrian-to-vehicle interactions in many different situations. To evaluate properly the latter, the use of small scenarios with a limited number of pedestrians and vehicles is not enough; we need experiments in a scenario as close as possible to our current cities: a highly populated urban area with cars and pedestrians.

In [21], the problem of the possibility of generating too many warnings to drivers is mentioned, but the evaluation does not deal with it and focuses instead on the reception of signals at an intersection, to see if, even in the presence of buildings, the reception is good enough to get the messages needed by the proposed warning system. In [29], a driver warning system to protect pedestrians using cellular communications is proposed. This work uses simulations with SUMO to evaluate the use of the warning system in an urban area. The proposed system warns of potential collisions between vehicles and pedestrians when, considering the speed of vehicles and pedestrians, it predicts that a vehicle and a pedestrian are going to end up in the same zone. False negatives and positives of the system are evaluated. False negatives happen when the system does not raise an alert, but the condition defined in the system to generate alerts finally occurs. In this system, this could happen because the prediction can be wrong, for example, because pedestrians can change their speed very quickly. False positives happen when the system raises an alert, but the condition defined in the system to generate alerts finally does not occur. Again, this can happen, for example, because of mistakes in predictions. The evaluation in [29] shows that there are some number of false negatives and positives in the proposed system. However, [29] does not provide information about the absolute number of alerts generated by the system or their length, and it does not assess the number of alerts that are raised correctly according to the warning system, but that do not result in a danger for pedestrians (unnecessary alerts). For example, the zone to raise an alert in the algorithm proposed in [29] covers the sidewalks in both sides of the road so, the system raises an alert when a car moves in one direction and there is a pedestrian in the sidewalk on the other side of the road. This is not a false positive, as it is what is expected from the design of the warning system, but it is an unnecessary alert because the pedestrian is never in danger. Our paper analyzes how this kind of alerts lead to an overload in drivers that jeopardize the feasibility of the pedestrian protection system.

**3. Algorithms for a communication-based warning system to protect pedestrians**

*3.1. The warning system*

Our work focuses on the use of a communication-based warning system to improve pedestrian safety. The system tries to warn drivers in situations where their vehicle can become a potential future risk to pedestrians, so drivers can increase their attention and be more cautious. Therefore, the system triggers an alert in a vehicle to warn the driver when some conditions are met. The conditions try to predict in advance that the vehicle might be a risk to a pedestrian. In the system, a vehicle has a different alert for each pedestrian that can be potentially at risk. An alert will be active as long as the conditions are still met. As proposed also in [21], the user interface in the vehicle may choose to aggregate simultaneous alerts to simplify the presentation of the warnings to drivers. We analyze the effect of this strategy in section 5, in which we evaluate, besides the absolute number of alerts, the amount of time in which at least one alert is active.

In this paper, we do not discuss detailed user interface aspects. Alerts may be represented by some specific sounds or lights or, in advanced systems, information may be projected on the windshield. Better user interface solutions can help drivers deal with a larger number of alerts, but even so, alerts are inputs that drivers must process, so warning systems have to limit these inputs whenever possible to avoid unnecessarily distracting drivers.

We have designed four different alternative algorithms for the warning system. Each algorithm represents a different strategy to trigger alerts, so we can compare them. An important requirement for the algorithms is to ensure the feasibility of their implementation in the real world. That is why our proposed algorithms only manage simple information that can be readily available with current technology. This information is both local and remote. Local information, available in the vehicle, is its location and heading. We also assume that vehicles have a digital map with the location of the crossings in the area. It would also be possible to equip the crossings with a device advertising its location (in fact, we use this approach in our experiments in Section 4). Remote information, not directly available in the vehicle, is the location of pedestrians. Therefore, we also need some communication mechanism to send this information from pedestrians to vehicles.

In our system, we assume that pedestrians have a device that sends beacons that include its location. This could be a dedicated device or an application in a mobile phone. Beacons are received in vehicles. Although other options are possible, in our proposal the communication between user devices and vehicles is based on a direct short-range radio system using Wi-Fi/IEEE 802.11. When a beacon is received, the algorithm in use in the warning system evaluates the situation and, if required, triggers an alert. Once an alert is active, a timer is used to deactivate the alert. If the timer expires without receiving a new beacon that confirms, according to the respective algorithm, the alert situation, the alert is deactivated. The timer of an alert is restarted each time a beacon that confirms that alert is received.

*3.2. The algorithms to decide when alerts are active*

The four algorithms are defined as follows:

1. Algorithm 0: This is the simplest approach. If a vehicle detects a pedestrian at a distance that is below a threshold, an alert is triggered for that pedestrian. The threshold is called alert distance threshold ($th_{ad}$), and is a configurable parameter in the algorithm: a larger threshold means more time to react but also more and longer alerts. The behavior of algorithm 0 is presented in Figure 1. In this figure, and the following, several example scenarios are shown, and those in which an alert is active are highlighted with a red circle, while those that do not activate an alert have a green circle. Note that Algorithm 0 is very simple and robust. It does not need to worry about the course of the vehicle, for example, as it only depends on the distance between pedestrians and vehicles. On the other hand, algorithm 0 can be expected to create many unnecessary alerts that distract the drivers. Therefore, algorithm 0 is intended as a baseline performance reference for our work.
2. Algorithm 1 (Figure 2): We add a restriction to Algorithm 0. In this case, an alert is triggered if the conditions of Algorithm 0 are met, but also the vehicle is close to a crossing. Therefore, an alert is active when a vehicle is close to a crossing and the distance between the vehicle and the pedestrian is below $th_{ad}$. The vehicle is considered close to a crossing using the same threshold as the one used to activate the alerts ($th_{ad}$). The reason for this is that in both cases the requirement is that a driver must be able to brake before traveling that distance.
3. Algorithm 2 (Figure 3): We add a new restriction to Algorithm 1. To trigger an alert, the vehicle not only has to be close to a crossing, the crossing must be in front of the vehicle. The crossing is considered in front of the vehicle if the crossing is located at an angle of ±90º around the direction of movement of the vehicle.
4. Algorithm 3 (Figure 4): We add a new restriction to Algorithm 2. To trigger an alert, the pedestrian that causes the alert must be in front of the vehicle and near the crossing. Therefore, an alert, associated with a pedestrian, is active when the vehicle is close to a crossing, the pedestrian is close to the same crossing, the pedestrian and the crossing are in front of the vehicle, and the distance between the vehicle and the pedestrian is below $th_{ad}$. The pedestrian is close to the crossing using a threshold that we call pedestrian safety threshold ($th_{ps}$).

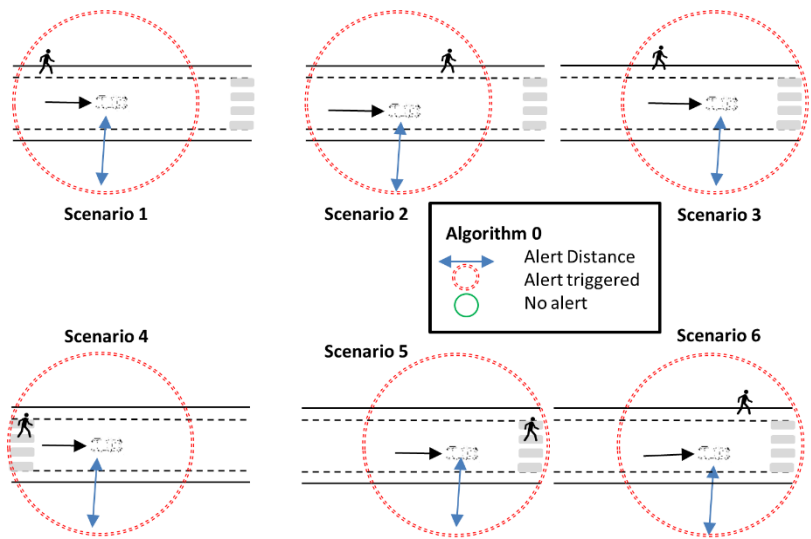

**Figure 1.** Alerts in algorithm 0

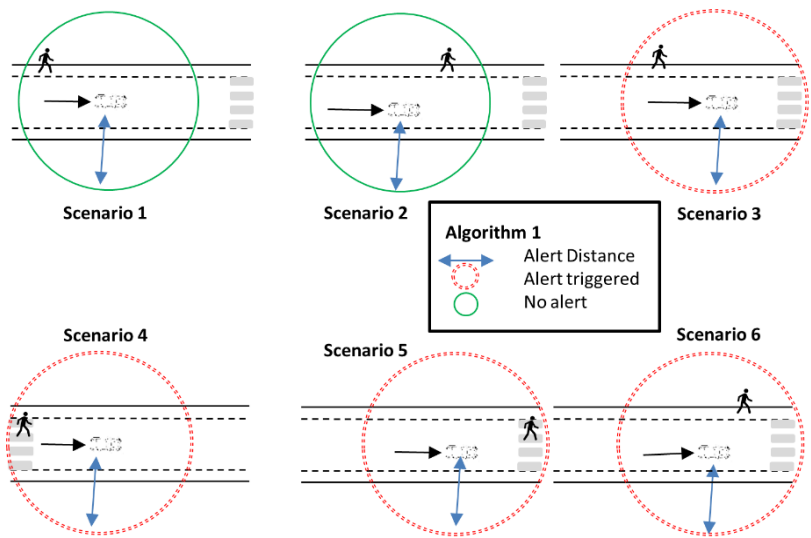

**Figure 2.** Alerts in algorithm 1

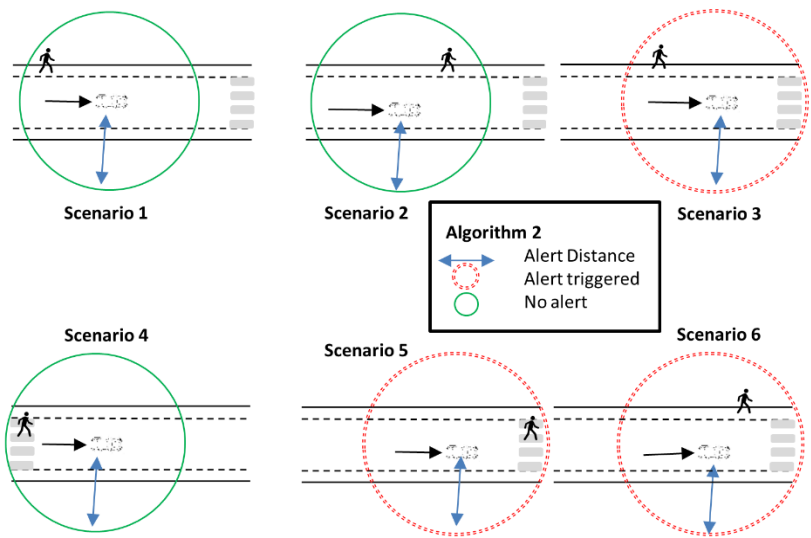

**Figure 3.** Alerts in algorithm 2

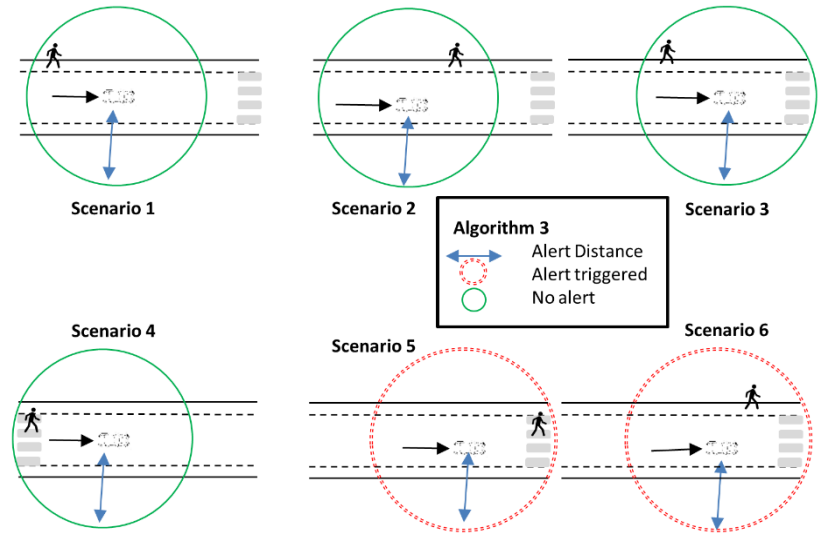

**Figure 4.** Alerts in algorithm 3

The four algorithms (see a summary in Table 1) represent improvements to better capture situations that can lead to a risk to pedestrians. We add restrictions that try to reduce the number of unnecessary alerts, i.e., triggered alerts that do not correspond to a real danger. For example, we focus on crossings, because warning drivers about any pedestrian walking on a sidewalk would create many unnecessary alerts. However, it is worthwhile to alert a driver approaching a crossing of the existence of a nearby pedestrian who could enter the crossing. Note that there could be some practical adaptations to this approach without changing the philosophy of the algorithms. Intersections could be considered as "crossings" by the algorithms even if they do not have a crossing, since they are places where pedestrians are likely to try to cross the road. Additionally, any pedestrian that leaves the sidewalk and enters the road could be immediately considered as "close to a crossing" by the algorithm, but this strategy would require more location accuracy than currently available on smartphones, although this could change in the future. Another strategy used in the algorithms is to focus on situations in front of the vehicle and not behind.

**Table 1:** Summary of algorithms

| Algorithm | Conditions |
|---|---|
| Algorithm 0 | $D_{Vehicle-pedestrian} < th_{ad}$ |
| Algorithm 1 | $D_{Vehicle-Pedestrian} < th_{ad}$<br>$D_{Vehicle-Crossing} < th_{ad}$ |
| Algorithm 2 | $D_{Vehicle-Pedestrian} < th_{ad}$<br>$D_{Vehicle-Crossing} < th_{ad}$<br>Crossing in front of vehicle |
| Algorithm 3 | $D_{Vehicle-Pedestrian} < th_{ad}$<br>$D_{Vehicle-Crossing} < th_{ad}$<br>Crossing in front of vehicle<br>Pedestrian in front of vehicle<br>$D_{Pedestrian-Crossing} < th_{ps}$ |

The number of alerts generated by an algorithm is of key importance for its feasibility. Ideally, we would like to trigger alerts only when a situation is going to lead to a risk for a pedestrian, and as early as possible. However, drivers will end ignoring an alert system that generates too many unnecessary alerts. Therefore, the warning system must capture risky situations while avoiding unnecessary alerts. To determine the best approach to create alerts to protect pedestrians, we can modify the distance between the vehicle and the pedestrian where we begin to trigger alerts ($th_{ad}$), and the type of algorithm (the intelligence of the algorithm in detecting potential risky situations).

In this paper, we will show that some of the algorithms do not achieve the expected improvements in reducing the number of unnecessary alerts. Therefore, it is important to understand that the design of these algorithms, which try to predict risky situations for pedestrians, is a complex issue, especially when considering the global effects as our research does.

*3.3. Configuration of the algorithms*

In the algorithms, we have two configurable parameters, the alert distance threshold ($th_{alert-distance}$) and the pedestrian safety threshold ($th_{ps}$, only for algorithm 3). We study next how to choose appropriate values for those parameters.

3.3.1. Alert distance threshold

In all the algorithms, there is no alert if the vehicle is far from pedestrians but, when the vehicle approaches a pedestrian, an alert can be triggered when the distance from the vehicle to the pedestrian reaches the $th_{ad}$. Therefore, under ideal conditions, an alert will be triggered at a distance from the vehicle to the pedestrian equal to $th_{ad}$. Ideally, to minimize the number of alerts, we would like $th_{ad}$ to be as small as possible. However, we also need $th_{ad}$ to be large enough to allow a vehicle to stop before traveling that distance. There is a rate of deceleration, *decel*, which allows the vehicle to stop at $th_{ad}$. We can relate $th_{ad}$, *decel*, and the initial speed of the vehicle, $s_{i_{veh}}$ by Equation (1):

$$th_{ad} = tr \times s_{i_{veh}} + \frac{0.5 \times (s_{i_{veh}})^2}{decel} \quad (1)$$

In Equation (1), the first addend accounts for the distance traveled before starting to brake due to the driver reaction time, $tr$; and the second corresponds to the distance traveled during braking.

We must choose a $th_{ad}$ that allows the vehicle to brake at a reasonable deceleration rate. Modern cars can achieve deceleration rates of more than 9 m/s² [33]. We select a more comfortable target deceleration rate of 5 m/s². We use 60 km/h (16.67 m/s) as the $s_{i_{veh}}$, which is higher than the legal limit for urban traffic in many cities, being a conservative choice. We assume that the driver reaction time is 0.5 s [34]. With these numbers, and according to Equation (1), the $th_{ad}$ must be greater than 36.11 m. In our simulation work, described in section 5, we explore different values of $th_{ad}$, namely 40 m, 70 m, and 100 m. This allows us to study the effect of providing different safety margins to drivers to deal with alerts.

3.3.2. Pedestrian safety threshold

In algorithm 3, this threshold, $th_{ps}$, defines the distance from the pedestrian to the crossing to trigger alerts. $th_{ps}$ must be large enough for a vehicle to react, avoiding putting the pedestrian at risk. The earliest time in which a danger is possible is after the pedestrian walks the distance to reach the crossing ($th_{ps}$). So, in the worst case, a vehicle must be able to stop in the time $t_{pc}$, which is the time the pedestrian needs to reach the crossing:

$$t_{pc} = \frac{th_{ps}}{s_{ped}} \quad (2)$$

where $s_{ped}$ is the speed of the pedestrian. The deceleration rate needed in the vehicle to stop in the time $t_{pc}$ is:

$$decel = \frac{s_{i_{veh}}}{t_{pc}-tr} \quad (3)$$

For example, if we want a *decel* ≤5 m/s²; with $s_{i_{veh}}$ = 60 km/h = 16.67 m/s; and $s_{ped} = 1.6$ m/s; applying Eqs. (3)-(2) results in that $th_{ps}$ must be greater than 6.13 m assuming a driver reaction time of 0.5 s [34]. We chose a conservative value of $th_{ps} = 10$ m for our simulations presented in section 5.

## 4. Evaluation of the feasibility of the defined alerts

In this section, we evaluate the feasibility of implementing the proposed algorithms in the real world. In the tests, the vehicles are fitted with physical communications equipment, and pedestrians are equipped with mobile phones with an installed application. The tests were carried out on the closed test tracks of the University Institute for Automobile Research (INSIA) of the Technical University of Madrid. The equipment used is the following:

1. Vehicular communication modules. These modules are based on an AR9220 chipset that uses the ath9k modified driver to allow the 802.11p bands. The operating system that incorporates the modules is a Debian wheezy with a 3.19.0 kernel version. The kernel has been configured to allow OCB (Outside the Context of a BSS) mode in the ath9k driver. This module also acts as a Wi-Fi router, which enables a 2.4 GHz network with neighboring devices (we have used this communication functionality in our experiments). Finally, the module includes the Global Navigation Satellite System NV08C-CSM chipset (NVC). It is an integrated GLONASS + GPS + GALILEO + SBAS satellite navigation receiver for use in various applications that demand low cost, low power consumption and uncompromised performance. One module is installed in the vehicle while another module is used to identify the position of the pedestrian crossing.
2. Motorola Nexus 6 smartphone. This phone is used to position the pedestrian and send this information to the vehicle. In particular, pedestrians use the smartphone to become part of the V2X environment. The smartphone oversees the execution of two actions. On the one hand, it uses its geo-position libraries to calculate its geographical coordinates. On the other hand, the smartphone connects automatically to the Wi-Fi network of the vehicle V2X on-board unit (OBU) when the phone is within its range of coverage. An ad-hoc application installed in the smartphone retrieves the position and sends it (in a custom message) to the vehicle (the OBU). In the vehicle, the pedestrian is added to the list of elements that navigates in the surroundings, and the pedestrian protection algorithms are applied.
3. Trimble® R4 DGPS receiver that provides an update rate of 10 Hz and works with differential correction via GPRS. Differential correction is performed by software NTRIP Client 2.0 and connects to a server of differential correction NTRIP (Networked Transport of RTCM via Internet Protocol) via a smartphone connection and bluetooth GPS equipment. Given its high accuracy, this equipment is used for the validation of the positions in the preliminary calibration studies and during the tests, in order to assess those cases in which the positioning errors could imply errors in the warnings (false positives or false negatives).

The tests performed are as follows:

1. Estimation of positioning errors (in static and dynamic conditions)
2. Tests for the analysis of the warning system

*4.1. Positioning errors estimation*

Although the accuracy of satellite positioning depends on the conditions of the site and when the measurements are taken, these tests are intended to have a rough estimate of the margins of error that can be considered as usual in the equipment used later in the analysis of the warning system. Since the warnings can be very dependent on the correct positioning, it is necessary to evaluate to what extent the algorithms' response can be influenced by the positioning errors.

Firstly, we analyzed the dispersion in the positioning of a static position with the 3 pieces of equipment used during a 2-hour test. The dispersion of the Trimble DGPS receiver is very low (0.02 m) in comparison with the other two (2.15 m and 3.27 m for the NVC GPS of the communication modules and the mobile phone, respectively) so it can be taken as reference. On the other hand, the mobile phone presents a somewhat different behavior than the other two receivers. Because the phone includes accelerometers, it can estimate whether it is moving or not so, when it does not detect movement, it offers a single fixed position for long intervals of time, which is only modified in certain instants. Then, it takes another position that is maintained again a certain time until the conditions of the satellites change significantly. This behavior differs from the oscillations of the other receivers, which fluctuate over time, without fixing a position due to the knowledge of being in a static location.

After that, we performed a test on the test track with a vehicle that equips three receivers in very close positions (two modules with NVC GPS and the Trimble DGPS receiver). The vehicle traveled following a path with continuous changes of direction, registering the differences in the positioning between the NVC GPS of the vehicle communication modules and the Trimble DGPS receiver (our reference). Figure 5 shows the registered positioning errors in the vehicle communication modules, errors that are bounded below 2 meters, after an initialization time. In this figure, and the following ones, the axes represent the recorded trajectories in Universal Transverse Mercator (UTM) coordinate system. Furthermore, we also verified the high repeatability in the response of the two communication modules, working under the same conditions, with a difference of less than 20 cm.

Finally, it should be noted that these results are dependent on the test conditions, but they can provide an idea of the order of magnitude of expected errors. In the following subsection, the performance of the proposed warning system is assessed, and the extent to which these positioning errors have influence on warnings is analyzed.

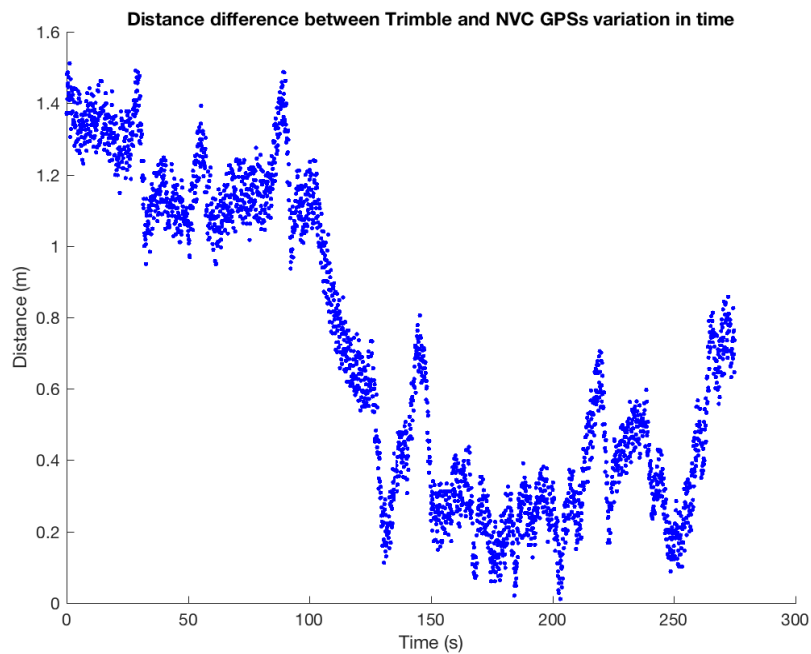

**Figure 5.** Differences in positioning

*4.2. Warning system analysis*

We have seen that it is on the mobile phone where we have found larger errors in positioning. Therefore, in this section, we analyze the effect on our proposed algorithms of positioning errors in the mobile phone. For this analysis, we address the six scenarios of Figures 1 to 4, which define key situations for the behavior of the algorithms of the warning system. To replicate the scenarios, we defined four test configurations involving a vehicle and a static pedestrian (see the testbed architecture in Figure 6). Table 2 shows the configurations used in the experiments. In each test, the vehicle follows a straight line, first arriving at the pedestrian location, afterwards at the crossing, and then leaving the crossing behind. Depending on the configuration, i.e., the distance between the pedestrian and the crossing, the activation and deactivation points of the alerts triggered by the different algorithms of our system are different. In these tests, we consider that $th_{ad}$ is 10 m. This threshold is much lower than the values that we would use in practice, but this fact simplifies the tests and, as we are interested in studying the impact on the system of positioning errors, this low $th_{ad}$ highlights the effects of those errors. More practical distances will be used in our simulations later to reproduce real potential scenarios, but these tests provide upper limits for errors in physical implementations.

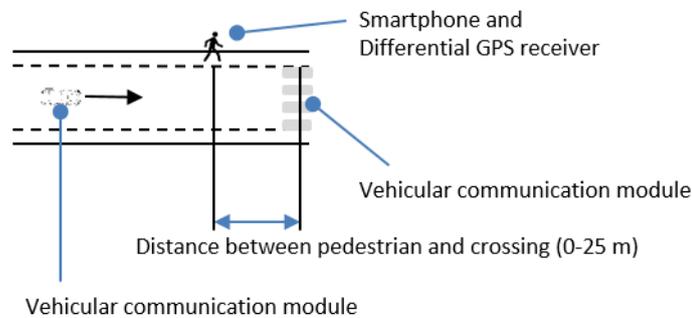

**Figure 6.** Testbed architecture

**Table 2:** Test configurations

| Configuration | Distance between pedestrian and crossing (m)[1] | Scenario defined in section 3 |
|---|---|---|
| 1 | 25 | 1-2 |
| 2 | 15 | 3 |
| 3 | 0 | 4-5 |
| 4 | 5 | 6 |

[1] The pedestrian is placed before the crossing in the vehicle movement direction

In all the tests, the crossing is located using a communication module (the same type as the one used in the vehicle). The position of the pedestrian is obtained using a mobile phone and the DGPS receiver, which allows us to compare the location provided by the mobile phone with an accurate reference (see Figure 6). For each configuration, we analyzed when each of the algorithms has alerts activated if the pedestrian is located using the mobile phone, as well as if the pedestrian is located using the DGPS receiver. Each test configuration is repeated 30 times. It should also be noted that, in the same test, by having the positions of the different GPS receivers, it is possible to determine the activations of alerts by each of the four algorithms. Figure 7 shows an example of the system operation. It displays the position of the crossing, the position of the pedestrian (provided by mobile phone and DGPS receiver), the vehicle path, and when the different types of alerts are active. In this figure, when an alert is indicated as active, all the alerts with lower number are also active. For example, the "Algorithm 3 alert" label implies that algorithms 2, 1, and 0 alerts are also active; and the change from label "Algorithm 3 alert" to "Algorithm 2 alert" implies that the alert of Algorithm 3 has been deactivated.

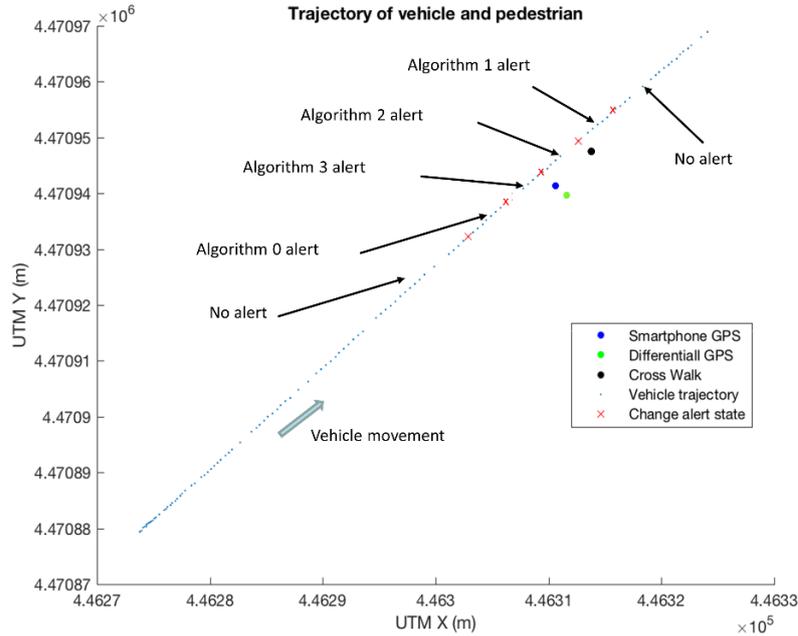

**Figure 7.** Example of warning system performance

Table 3 shows the results obtained in the set of 30 trials. We observe a high reliability of the algorithms, with low dependence on the errors of the mobile phone positioning. These errors (deviation DGPS-mobile phone in Table 3) are smaller than 2 m except in the case of configuration 2, where we recorded deviations greater than 5 meters. We evaluated how these deviations influence the alerts generated by each of the algorithms, comparing the results with the mobile phone and the reference provided by the DGPS receiver. Table 3 shows the differences detected in the amount of time alerts are active for the different algorithms and test configurations. Specifically, it shows, for each type of alert (algorithm): the total duration of alerts accumulated in all the tests in the same configuration, the sum of times when the alerts using the DGPS receiver and using the mobile phone are different (discrepancies in alerts), and the maximum time of this difference recorded for any of the tests.

We can see in Table 3 that, in test configuration 1, which includes scenarios 1 and 2 where only algorithm 0 activates alerts but not the others, we obtained the expected results. In scenario 3, no discrepancies were detected either.

**Table 3:** Results of the warning system performance under real positioning conditions

|  |  | Configuration | | | |
|---|---|---|---|---|---|
|  |  | 1 | 2 | 3 | 4 |
| **Deviation DGPS-mobile phone (m)** | Average value | 1.01 | 2.83 | 1.20 | 1.61 |
|  | Standard deviation | 0.37 | 1.01 | 0.31 | 0.29 |
| **Total duration of alerts (s)** | Algorithm 0 | 33.15 | 30.30 | 30.15 | 27.45 |
|  | Algorithm 1 | 0 | 17.70 | 26.85 | 24.30 |
|  | Algorithm 2 | 0 | 14.55 | 9.45 | 7.65 |
|  | Algorithm 3 | 0 | 2.55 | 9.45 | 7.35 |
| **Discrepancies in alerts** | Total (s) | 0 | 2.73 | 0 | 0.75 |
|  | Total (%) | 0 | 4.19 | 0 | 1.12 |
|  | Maximum (s) | 0 | 0.22 | 0 | 0.14 |

In test configuration 2, the deviations in the positioning were higher than in the other cases, as previously discussed. In this scenario, alerts with algorithms 0, 1 and 2, but not with 3, should have been recorded. However, we observe that alerts with algorithm 3 are activated for a total of 2.55 s (false positive). In any case, the discrepancies in the duration of the alerts do not exceed 1 second in this scenario, despite the high error in the positioning, the longest discrepancy was 0.22 s for a specific test. Figure 8 shows one of the trials in which a false positive occurs. The figure shows the trajectory of the vehicle, and the position of the pedestrian and the crossing. As it can be appreciated, the positioning errors for the smartphone are very high so that, although the configuration is according to case 2, in which the pedestrian is 15 m from the crossing, the data provided by the phone resembles more configuration 4, with the pedestrian at 5 m. In this way, with an accurate positioning, algorithm 3 would not raise an alert in this test, since the pedestrian and the crossing cannot be simultaneously located less than 10 m away from the vehicle and in front of it at any moment. However, the positioning error for the smartphone causes the system to consider that both elements are in front of the vehicle in a radius of 10 m, so the alert is triggered.

Finally, in test configuration 4, discrepancies in the alerts have been recorded for 0.75 s, representing 1.12% of their total duration. Figure 9 shows the differences in alert generation between the smartphone GPS and the DGPS for a trial of configuration 4, in which the duration of the alert of algorithm 3 considering the positioning of the smartphone is longer than with the positioning of the DGPS receiver. However, the discrepancy is very small and negligible in a practical application because a driver in practice cannot notice the maximum discrepancies.

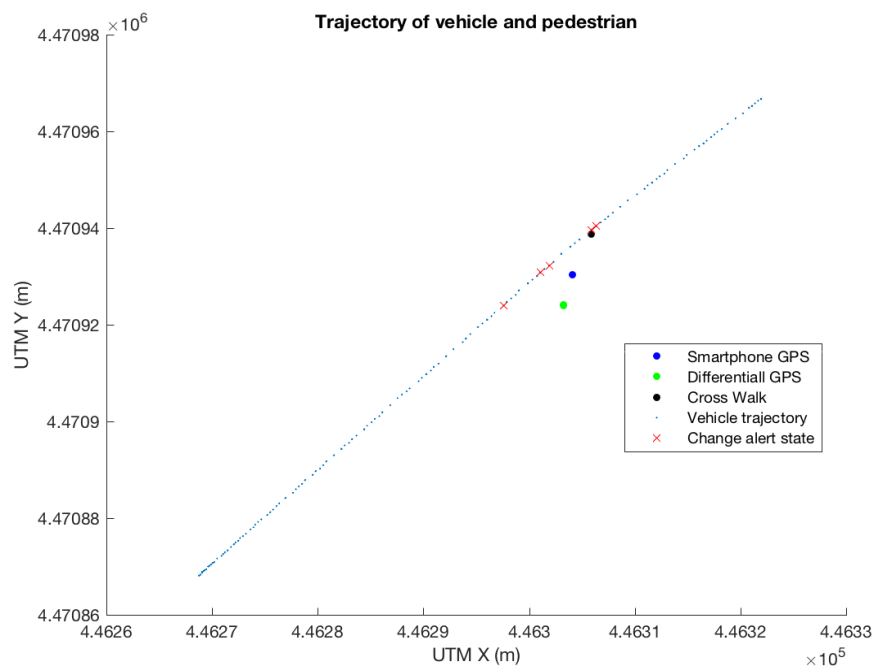

**Figure 8.** Example of warning discrepancies in test configuration 2. Pedestrian, crossing and vehicle positioning;

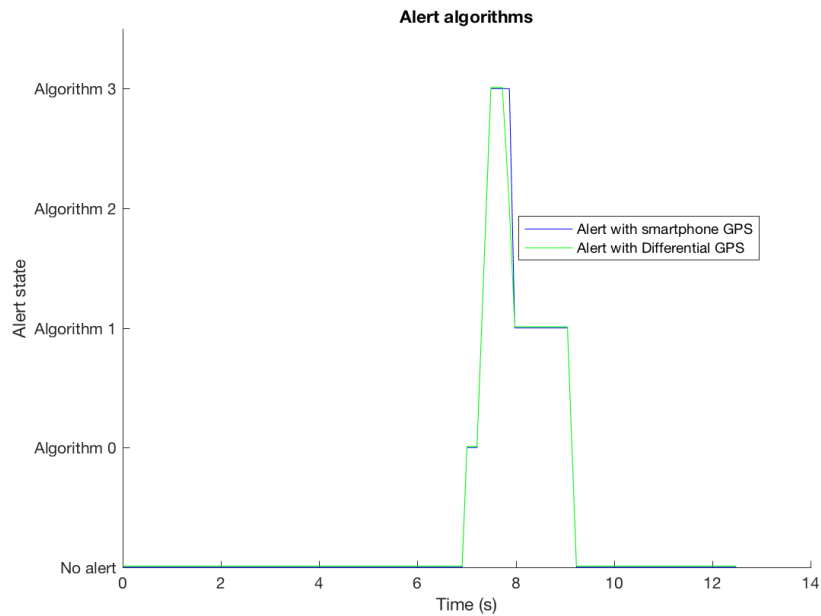

**Figure 9.** Example of warning discrepancies in test configuration 4.

In the tests we have found that, despite the positioning errors, the system generates a low percentage of false positives (i.e., instances in which the system triggers an alert but the conditions defined in the algorithm to raise alerts have not really happened). Besides, it does not have false negatives (i.e., there are no instances in which the system omits alerts when it should have triggered one). Even in a set of tests in which recorded positioning errors have been quite high (higher than what static and dynamic positioning tests have provided previously), the warnings are only slightly influenced. Therefore, we can conclude that the defined algorithms are robust against positioning inaccuracies, and their practical implementation is viable even on basic equipment such as mobile phones and communication modules in vehicles.

## 5. Evaluation of the behavior of the algorithms using simulation

In this section, we analyze the behavior of our proposed four algorithms. We focus on the number and duration of alerts that are generated, which is related to the mental load that the warning system produces on drivers. These aspects must be studied in a large-scale deployment to accurately represent the interactions of many vehicles and pedestrians as they travel through a city. Large-scale deployments in the real world are difficult and expensive, and it is difficult to achieve proper conditions at any time, so a good approach is to use simulations. However, to guarantee the validity of the results, it is of crucial importance to use realistic traffic simulators.

*5.1. Simulation tool*

For representing a realistic urban scenario, we have chosen the SUMO (Simulation of Urban MObility) road traffic simulator [35,36]. SUMO provides an accurate modeling of microscopic vehicular traffic, simulating the behavior of pedestrians and vehicles. SUMO provides several models, with different parameters, for representing vehicles' behavior. Our simulations use the default models and parameters in SUMO version 0.30.0. The key components are the car-following model, which is the SUMO extended Krauss model [36]; and the lane changing model, which is the one described in [37]. The interaction of vehicles and pedestrians at crossings is regulated by the SUMO pedestrians' model. In this model, using what is called crossings without priority in SUMO, pedestrians wait to cross if they do not have enough time to do so before any approaching vehicle reaches the crossing, but vehicles stop at crossings if any pedestrian is crossing.

In our simulations, vehicles are controlled by SUMO. They do not react to the alerts; SUMO by itself avoids accidents although it can be quite aggressive allowing vehicles to get very close to pedestrians. The aim is to study if the warning system helps to detect situations that represent a real risk for pedestrians. For this reason, in our simulations, we do not want to use alerts to stop vehicles. In this way, we can evaluate whether each alert was correct (i.e., necessary) or not. For the same reason, we have chosen to use crossings without priority for pedestrians, so SUMO does not try to stop vehicles to allow pedestrians to cross. In our SUMO simulations, it is the pedestrian who stops to avoid a collision (to allow the vehicle to pass) but, if the pedestrian has to stop, our evaluation will detect this as a risk situation (there would have been an accident if the pedestrian had not stopped).

The other key component of our simulation framework is MASON [38,39]. MASON is a multi-agent simulation toolkit that allows efficiently implementing applications with many individual entities that interact with each other. We use MASON to control the SUMO simulation, which means that MASON starts and decides when to advance each step of the SUMO simulation. Each vehicle and pedestrian in SUMO is represented by an agent in MASON. SUMO provides the behavior of vehicles and pedestrians from the point of view of traffic interactions. MASON, in each step of the simulation (100ms of time in the simulation), gets the current state from SUMO, with all the information about vehicles and pedestrians. The sending and reception of beacons, processing of the alert algorithms, and statistics collection are done in MASON.

The integration of SUMO and MASON is done using TraCI (Traffic Control Interface), provided by SUMO. TraCI is a protocol that allows access to a SUMO running simulation. We have used TraCI4J [40], a Java implementation of TraCI (MASON is a Java program). TraCI also allows the external program, MASON in our case, to send commands at any point in time to control individual vehicles. MASON, using TraCI, can instruct vehicles in SUMO to react to alerts (for example, by braking). However, as explained above, in our simulations, vehicles do not react to alerts and, therefore, we have not used this TraCI functionality.

*5.2. Urban area map*

The simulations take place in an urban area (800 m x 700 m) of the city of Madrid (Spain). The map is shown in Figure 10. The size of the urban area, and the presence of different types of streets, allows us to study multiple interactions between vehicles and pedestrians in diverse situations. The map was obtained from OpenStreetMap (OSM) [41]. We have used JOSM [42], a Java editor for OpenStreetMap, in order to extract the desired area, and to verify the junctions and connections. Finally, using the NETCONVERT tool of SUMO, the map is converted to the XML format that SUMO requires (it contains the information about roads, traffic signs, junctions, crossings…).

The legal speed limit of the roads of the map is 50 km/h. The maximum speed of a vehicle in the simulations is obtained from a normal distribution centered on the speed limit, deviation equal to 0.1, and capped at 20% and 200% of the speed limit. This means that, in any simulation, 95.44% of the vehicles have a maximum speed between 40 km/h and 60 km/h, and no vehicle has a maximum speed below 10 km/h or above 100 km/h. Besides, vehicles try to adapt their speeds to the traffic conditions according to the vehicle model in SUMO (e.g., reducing the speed when approaching an intersection or when the vehicle in front is slower and overtaking is not possible).

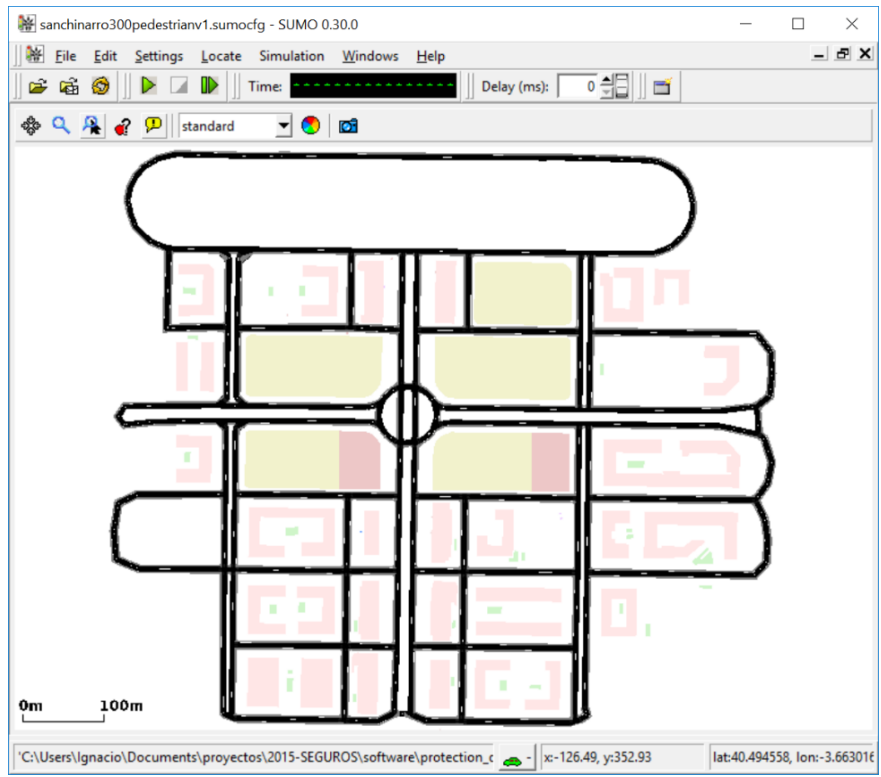

**Figure 10.** Map of the scenario (SUMO user interface)

The target speed of pedestrians is 5.76 km/h (1.6 m/s). This value is consistent with [43], a study that measured the walking speed in different countries, and found that it was between 3.93 km/h and 5.92 km/h. In SUMO, pedestrians try to walk at the target speed, but they adapt their movement to the traffic conditions, and they also decelerate randomly, at each step, a maximum of 0.2 of the target speed. This is according to the SUMO pedestrian model, and it achieves a natural behavior in which pedestrians walk at a varying speed close to the target speed. Pedestrians walk along the sidewalks available on all the streets on the map. Pedestrians can only cross the roads using the pedestrian crossings available at every intersection. As mentioned above, none of the crossings on the map has traffic lights to give priority to pedestrians.

We also need information about buildings in the scenario, since buildings can interfere with the reception of beacons sent by pedestrians. We generated the buildings manually. We used a custom developed tool, and created a building filling each block of the map. This is a harder environment than the reality, where there are no buildings in some of the blocks, or the buildings do not cover all the surface of a block between roads. Figure 11 presents the map as seen by MASON, with the information of buildings (in blue) and crossings (small red lines in intersections). The borders of the map are imported from SUMO via TraCI. The coordinate system used in SUMO is mapped to the map in MASON.

*5.3. Traffic conditions*

Using SUMO tools, we have defined some vehicular and pedestrian traffic. The arrival processes are based on a binomial distribution. During an hour, each second an arrival happens with probability $\frac{1}{\text{inter arrival time}}$. The inter arrival time of vehicles entering the simulation is 7.2 s (therefore, on average, 500 vehicles participate in each simulation). Each vehicle has a trip associated, i.e., it has to go from one random point in the map to another (separated by at least 600 m). Pedestrians also enter the scenario during an hour. We explore different pedestrian densities: inter arrival times of 12s (300 pedestrians, on average, in the simulation), 7.2 s (500 pedestrians), and 5.13 s (700 pedestrians).

Each pedestrian also has a trip associated, i.e., she/he has to go from one random point in the map to another separated at most 1000 m.

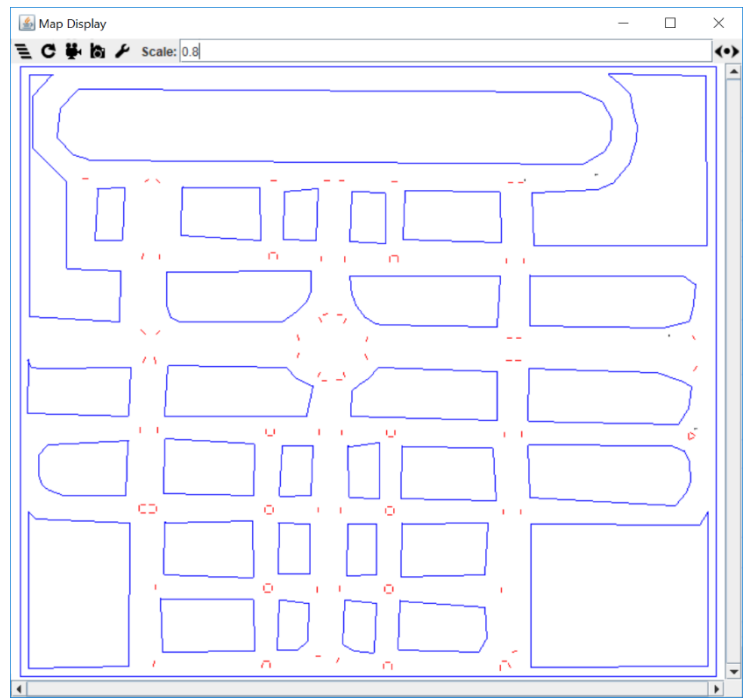

**Figure 11.** Representation of the map in MASON

The routes, i.e., how to go from the origin to the destination defined in the trips, are calculated using the DUAROUTER tool provided by SUMO. It applies the Dynamic User Assignment (DUA) algorithm [44,45] to calculate the routes in such a way that it is not possible to reduce the travel cost of any of the routes by going through another path. The DUA algorithm calculates these routes using an iterative process that we limit to 30 repetitions. With this process, we obtain the starting time, the starting position in the map, and the intended route for each vehicle and pedestrian participating in the simulations[1]. During the simulations, to travel the routes, SUMO applies the microscopic traffic model, realistically simulating the behavior of vehicles and pedestrians.

To characterize the traffic in our simulations, we have calculated the average stay time of a vehicle in the scenario, which is 134.92 s, and the average stay time of a pedestrian, which is 344.82 s.

### 5.4. Beacon system

Our four algorithms for the warning system (Table **1**) are based on the use of beacons sent by pedestrians to announce their locations. In our simulations, pedestrians send beacons each 300 ms. This time period must be low enough to allow a reasonable update of the pedestrians' position, so pedestrians should not be able to walk a significant distance in that time period.

We do not use a radio propagation model, instead we limit the range of beacons to 100 m and direct line of sight. Therefore, beacons are received in vehicles within 100 m of the pedestrians sending them, but only if there is a direct line of sight between the pedestrian and the vehicle. This means that, in our simulations, beacons do not travel through buildings although, in reality, buildings decrease the range of propagation of beacons and this does not necessarily translate into a lost beacon.

---

[1] The SUMO map, the SUMO configuration file for each simulation, the trips, and the routes are available at http://www.it.uc3m.es/isoto/protection_of_VRU.zip. We hope that other researchers might be able to use the scenarios as a test case to compare different systems to protect VRUs.

The timer that we use to deactivate alerts is initialized to 1 s. Therefore, after 1 s without receiving a beacon confirming an alert, the alert is deactivated. This value protects the system for wrongly deactivating alerts when, for example, beacons are lost. In our configuration, with 300ms between beacons, with 1 s we give time to receive three beacons updating the location of the pedestrian before deactivating an alert (we could lose two consecutive beacons). The timer also means that alerts remain active longer than it would be needed, while the system confirms that the alert is not valid anymore, but this additional time is negligible.

*5.5. Configuration of the simulations*

For each pedestrian density (300, 500, and 700 pedestrians), we have five different sets of trips of pedestrians and vehicles. This allows us to have five different simulations for each scenario, to get averages and 95% confidence intervals.

We have evaluated each algorithm defined in Section 3. The other parameter that we have explored is the threshold of the distance between a vehicle and a pedestrian to trigger alerts (the alert distance threshold, $th_{ad}$). As discussed in section 3.3, we have considered three different values for this threshold: 100 m, 70 m, and 40 m. With lower thresholds, we have a lower number of alerts but also less time to react to a potential risk situation.

In algorithm 3, we also have the pedestrian safety threshold, $th_{ps}$, which we have set equal to 10 m, as discussed in section 3.3.

*5.6. Simulation results*

To validate the correct behavior of the warning algorithms, we first needed to verify that they create alerts for any situation that results in a danger to a pedestrian, i.e., that the filtered alerts are really unnecessary alerts. We have tested this by defining, in our simulations, a danger situation as one in which a pedestrian is at a crossing or within 1m of the endpoints of it, and a vehicle is close and approaching (the distance between the pedestrian and the vehicle is less than 5 m and it is decreasing). We counted the number of times a danger situation happened (on average, we had 174 danger situations in the simulations with 300 pedestrians, 285 in those with 500 pedestrians, and 379 in those with 700 pedestrians). These numbers were measured using real distances in the simulator to avoid potential interference due to lost beacons (so, it is the real number of danger situations). Then, for each danger situation detected, we checked, for each algorithm, if an alert was active for that danger situation. All the algorithms achieved a 100% detection rate of danger situations, i.e., each algorithm had an active alert for each danger situation. Therefore, the algorithms do not have false negatives.

Once we have verified that the algorithms do not lose real danger situations, we analyze their performance. First, we look at the average number of alerts triggered in each vehicle during its travel (Table 4). An excessive number of alerts means overloading the driver, so the alerts lose their usefulness. As we can see in Table 4, an appropriate algorithm can result in a significant saving in the number of alerts. More specifically, algorithm 3 triggers at least 50% fewer alerts than the other algorithms for the same alert distance threshold. It is interesting also that algorithms 1 and 2 fail to reduce the number of alerts, and algorithm 2 even increases it in many cases. We tracked the reason in the simulations and, to show it, we describe next an example of a typical situation of interaction vehicle-pedestrian. A pedestrian is at half distance between two crossings (at a distance shorter than $th_{ad}$ from each of them). The two crossings are separated by a distance longer than $th_{ad}$. A vehicle moves from one place ahead of the first crossing to another behind the second crossing. In this scenario, algorithm 0 will have just one alert (around the location of the pedestrian). Algorithm 1 will also have one alert (as the two crossings are separated by a distance shorter than two times $th_{ad}$, the vehicle has one crossing or the other within a distance equal to $th_{ad}$). Algorithm 2 will have two alerts (the first one finishes after passing the first crossing, and the new one activates when reaching a distance to the second crossing equal to the $th_{ad}$). Finally, algorithm 3 will have none (if the pedestrian is not within a distance equal to $th_{ps}$ of any of the crossings) or just one.

Even more interesting than the number of alerts is the amount of time that a vehicle is in an alert condition during the simulation. This is shown in Table 5. We define an alert condition as a situation in which the vehicle has one or more active alerts. Therefore, the amount of time in alert condition is the amount of time in which the vehicle has at least one active alert. These times must be compared with the total time a vehicle spends in the simulation (on average 134.92 s). We can see once again that a good algorithm can reduce the amount of time drivers are subjected to the stress of alert signals and, in particular, algorithm 3 achieves a significantly shorter amount of time in alert condition compared with the other algorithms. Note also, that even if algorithm 2 results in a slight increase in number of alerts compared to algorithm 1, it reduces the time in alert condition. Going back to the example, algorithm 2 creates two alerts for one alert in algorithm 0, but the time in alert condition is reduced as the first alert ends after passing the first crossing, while in algorithm 0 the alert remains active.

Summarizing, algorithm 3 achieves important savings in both the number of alerts triggered and the amount of time in alert condition. Algorithms 1 and 2, even being more elaborate than algorithm 0, fail to achieve the expected improvements. The important point here is that algorithms with good properties, when evaluated locally, can result in not–so-good behavior in the multiple interactions that happen while travelling through a city.

Finally, we must test if the algorithms, and more specifically algorithm 3, still warn drivers with enough time to react to a dangerous situation. In Table 6 we show the average distance between a vehicle and a pedestrian at which alerts are triggered (the alert trigger distance) for the different algorithms and the different $th_{ad}$. Comparing the alert trigger distance with the $th_{ad}$ is a good measure of the effect of the algorithm on the time provided to the driver to deal with the alert. The maximum possible alert trigger distance is $th_{ad}$, because the different algorithms never trigger an alert when the distance between the vehicle and the pedestrian is longer than the threshold. Therefore, ideally, we would like the alert trigger distance to be equal to $th_{ad}$. But the alert trigger distance ends up being shorter than the threshold due to lost beacons, to conditions in the algorithms that do not allow an alert even when the distance between the vehicle and the pedestrian is lower than $th_{ad}$, and to the times when vehicles or pedestrians enter the scenario.

We can see that algorithm 3 achieves a good trigger distance in alerts. It is usually close to $th_{ad}$ (between 82% and 90% of the threshold) and it is comparable to, or better than the alert trigger distance of the other algorithms. Only algorithm 0 achieves longer trigger distances in alerts for all the scenarios. But note that it is better (in terms of fewer alerts generated, less time in alert condition, and longer trigger distances) to use algorithm 3 with a longer $th_{ad}$ (e.g., 100m) than to decrease the $th_{ad}$ of algorithm 0 (e.g., to 70 m or even 40 m).

**Table 4:** Average and Confidence Intervals (95%) of the number of alerts triggered per vehicle for different pedestrian densities and $th_{ad}$

|  | 300 pedestrians | | | | | | 500 pedestrians | | | | | | 700 pedestrians | | | | | |
| --- | --- | --- | --- | --- | --- | --- | --- | --- | --- | --- | --- | --- | --- | --- | --- | --- | --- | --- |
|  | $th_{ad}$=100m | | $th_{ad}$=70m | | $th_{ad}$=40m | | $th_{ad}$=100m | | $th_{ad}$=70m | | $th_{ad}$=40m | | $th_{ad}$=100m | | $th_{ad}$=70m | | $th_{ad}$=40m | |
|  | Avg. | CI | Avg. | CI | Avg. | CI | Avg. | CI | Avg. | CI | Avg. | CI | Avg. | CI | Avg. | CI | Avg. | CI |
| **Algorithm 0** | 16,07 | ±1,05 | 13,24 | ±0,81 | 10,05 | ±0,65 | 26,11 | ±1,62 | 21,46 | ±1,26 | 16,11 | ±0,94 | 35,94 | ±2,63 | 29,54 | ±2,24 | 22,09 | ±1,91 |
| **Algorithm 1** | 16,09 | ±1,04 | 14,89 | ±0,88 | 9,56 | ±0,61 | 26,07 | ±1,69 | 23,99 | ±1,48 | 15,33 | ±0,93 | 35,96 | ±2,61 | 33,03 | ±2,74 | 20,97 | ±1,72 |
| **Algorithm 2** | 18,30 | ±1,20 | 15,87 | ±0,99 | 10,08 | ±0,62 | 29,77 | ±1,78 | 25,72 | ±1,57 | 16,26 | ±1,04 | 40,97 | ±3,31 | 35,25 | ±2,56 | 22,38 | ±1,67 |
| **Algorithm 3** | 7,22 | ±0,35 | 5,53 | ±0,28 | 3,95 | ±0,25 | 11,83 | ±0,89 | 9,08 | ±0,63 | 6,44 | ±0,51 | 16,34 | ±1,29 | 12,53 | ±0,96 | 8,84 | ±0,67 |

**Table 5:** Average and Confidence Intervals (95%) of the average time (in seconds) spent per vehicle in alert condition for different pedestrian densities and $th_{ad}$

|  | 300 pedestrians | | | | | | 500 pedestrians | | | | | | 700 pedestrians | | | | | |
| --- | --- | --- | --- | --- | --- | --- | --- | --- | --- | --- | --- | --- | --- | --- | --- | --- | --- | --- |
|  | $th_{ad}$=100m | | $th_{ad}$=70m | | $th_{ad}$=40m | | $th_{ad}$=100m | | $th_{ad}$=70m | | $th_{ad}$=40m | | $th_{ad}$=100m | | $th_{ad}$=70m | | $th_{ad}$=40m | |
|  | Avg. | CI | Avg. | CI | Avg. | CI | Avg. | CI | Avg. | CI | Avg. | CI | Avg. | CI | Avg. | CI | Avg. | CI |
| **Algorithm 0** | 77,03 | ±2,90 | 65,67 | ±2,55 | 45,71 | ±1,89 | 88,16 | ±3,99 | 77,81 | ±3,98 | 58,52 | ±3,30 | 96,22 | ±3,48 | 86,96 | ±3,44 | 68,58 | ±3,12 |
| **Algorithm 1** | 75,49 | ±2,46 | 61,44 | ±1,87 | 38,26 | ±1,27 | 85,06 | ±3,68 | 71,86 | ±3,38 | 48,62 | ±2,27 | 92,38 | ±2,56 | 79,60 | ±2,31 | 56,34 | ±1,64 |
| **Algorithm 2** | 63,89 | ±1,64 | 49,85 | ±1,32 | 28,67 | ±0,92 | 71,47 | ±2,70 | 57,85 | ±2,37 | 36,09 | ±1,50 | 77,13 | ±1,86 | 63,65 | ±1,45 | 41,45 | ±1,17 |
| **Algorithm 3** | 28,83 | ±1,05 | 21,22 | ±0,86 | 12,89 | ±0,65 | 38,03 | ±2,12 | 28,93 | ±1,62 | 18,53 | ±1,19 | 44,98 | ±1,45 | 34,84 | ±1,30 | 23,08 | ±1,03 |

**Table 6:** Average and Confidence Intervals (95%) of the vehicle-pedestrian distance (in meters) when an alert is triggered for different pedestrian densities and $th_{ad}$

|  | 300 pedestrians | | | | | | 500 pedestrians | | | | | | 700 pedestrians | | | | | |
| --- | --- | --- | --- | --- | --- | --- | --- | --- | --- | --- | --- | --- | --- | --- | --- | --- | --- | --- |
|  | $th_{ad}$=100m | | $th_{ad}$=70m | | $th_{ad}$=40m | | $th_{ad}$=100m | | $th_{ad}$=70m | | $th_{ad}$=40m | | $th_{ad}$=100m | | $th_{ad}$=70m | | $th_{ad}$=40m | |
|  | Avg. | CI | Avg. | CI | Avg. | CI | Avg. | CI | Avg. | CI | Avg. | CI | Avg. | CI | Avg. | CI | Avg. | CI |
| **Algorithm 0** | 87,72 | ±0,22 | 65,78 | ±0,11 | 38,69 | ±0,08 | 87,89 | ±0,21 | 65,95 | ±0,11 | 38,77 | ±0,05 | 87,82 | ±0,24 | 65,93 | ±0,13 | 38,77 | ±0,05 |
| **Algorithm 1** | 87,38 | ±0,26 | 61,11 | ±0,34 | 36,25 | ±0,28 | 87,56 | ±0,24 | 61,38 | ±0,18 | 36,35 | ±0,10 | 87,44 | ±0,24 | 61,40 | ±0,17 | 36,37 | ±0,02 |
| **Algorithm 2** | 80,02 | ±0,64 | 57,05 | ±0,20 | 32,65 | ±0,20 | 80,38 | ±0,48 | 57,36 | ±0,27 | 32,81 | ±0,28 | 80,34 | ±0,40 | 57,44 | ±0,06 | 32,82 | ±0,06 |
| **Algorithm 3** | 82,33 | ±0,59 | 62,22 | ±0,33 | 36,56 | ±0,21 | 82,22 | ±0,40 | 62,33 | ±0,24 | 36,62 | ±0,13 | 82,30 | ±0,25 | 62,37 | ±0,23 | 36,63 | ±0,14 |

To perform a more detailed analysis of how algorithm 3 behaves in terms of providing sufficient margin to drivers to avoid accidents, now we study the decelerations needed to stop vehicles in time to avoid accidents for each alert of algorithm 3. This analysis is similar to the reasoning done in section 3.3 to configure the thresholds used in the algorithms, but now we use real distances and speeds measured in the simulations instead of the theoretical objectives utilized to define the thresholds. The aim of this analysis is to verify that, for all the alerts, the deceleration needed to avoid an accident is within the limits of what can be achieved with current cars. Modern cars can achieve deceleration rates of more than 9 m/s² [33] therefore we use this value as a reference. Algorithm 3 controls two different aspects to guarantee the protection of a pedestrian. One is the distance between the vehicle and the pedestrian, and the other is the distance between the pedestrian and the crossing. If a pedestrian is far from a crossing, it does not really matter if the vehicle is close to the pedestrian, there is no risk until the pedestrian reaches the crossing. Therefore, for a particular alert, the deceleration needed to avoid an accident, *decel*, is the minimum of two values: the deceleration needed to stop the vehicle before reaching the pedestrian location, and the deceleration needed to stop the vehicle before the pedestrian reaches the crossing. In both cases, we consider a reaction time of the driver to the alert of 0.5 s [34]. Therefore:

$$decel = min\left\{\frac{0.5 \times (s_{i_{veh}})^2}{d_{vp} - tr \times s_{i_{veh}}}, \frac{s_{i_{veh}}}{t_{pc} - tr}\right\} \qquad (4)$$

where,

$$t_{pc} = \frac{d_{pc}}{s_{ped}} \qquad (5)$$

and $s_{i_{veh}}$ is the initial speed of the vehicle when the alert is triggered (measured in the simulation); $d_{vp}$ is the distance between the vehicle and the pedestrian when the alert is triggered (measured in the simulation); $t_{pc}$ is the time that the pedestrian needs to walk the distance $d_{pc}$; $d_{pc}$ is the distance between the pedestrian and the crossing when the alert is triggered (also measured in the simulation); $s_{ped}$ is the speed of the pedestrian; and $tr$ is the driver reaction time to the alert. For $s_{ped}$ we use the expected maximum speed of pedestrians (1.6m/s) instead of the actual speed of the pedestrian measured in the simulation when the alert is triggered, because the speed of the pedestrian can increase after the alert, given that the pedestrian may not be aware of the danger of an accident.

In this analysis, we ignore the alerts triggered just when a vehicle or pedestrian enters the simulation, as this can happen at any distance from pedestrians to vehicles. In

Figure 12, we show, for any other alert triggered by algorithm 3 in the five simulations with 700 pedestrians and $th_{ad}$ equal to 40 m, the deceleration that would be needed to avoid an accident. In the worst case, the deceleration is 6.03 m/s², which is within the braking possibilities of modern cars.

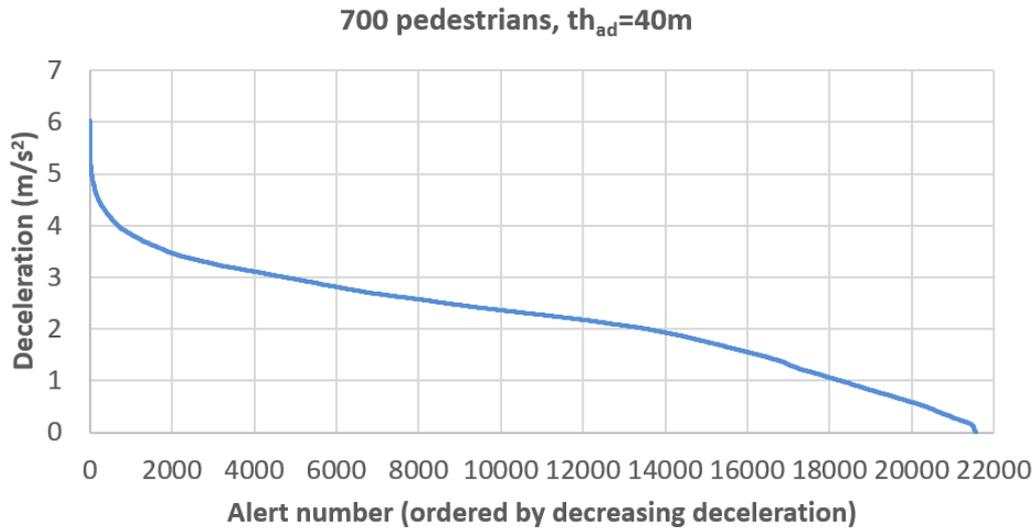

Figure 12. **Deceleration required for avoiding accidents after alerts (algorithm 3)**

*5.7. Discussion*

Algorithm 3 bases the decision to trigger an alert on the distance between the vehicle and the pedestrian, but considering only situations that happen in front of the vehicle and in places where pedestrians can cross the road. We have shown that the algorithm does a good job of identifying situations in which a vehicle can become a danger to a pedestrian, that the alerts are generated with enough time to deal with the situation in a safe way, and that the load generated in drivers with unnecessary alerts is significantly reduced compared with other algorithms.

Although the behavior of algorithm 3 achieves our initial objectives, there are potential improvements that we plan to analyze in our future research. The weakest property of the algorithm is the control of the worst case of the safety margin with which alerts are activated (measured as distance, or deceleration needed to stop the vehicle in time). The alert distance threshold, $th_{ad}$, controls the average safety margin and influences the worst case. But due to the effect of the angle check to generate the alerts, which is used to consider only what happens in front of the vehicle, it is possible that certain maneuvers (e.g., a U-turn) could result in situations in which an alert is triggered at distances below $th_{ad}$. We have seen this effect in the simulations, in which the worst case requires a deceleration rate to stop the vehicle in time that is still reasonable, but it is bigger than the one for which the $th_{ad}$ is configured. Note that the kind of maneuvers that lead to these situations imply a major change of direction, so they are low-speed maneuvers, which improves the safety margin even with lower distances. So, as we have seen in the simulations, they have not created practical problems, i.e., the worst-case safety margin is still good enough. However, to offer guarantees, we would need some additional mechanism in the algorithm. For example, we could add a mechanism to inhibit the angle check when the pedestrian is too close to the vehicle (some distance below the alert distance threshold, but enough to guarantee some worst case of the safety margin). On the downside, a mechanism like this one would also imply more and longer alerts.

We could also enhance the algorithm by considering the speed of the vehicles to define a dynamic $th_{ad}$. This could further improve the filtering of alerts (reducing unnecessary alerts), allowing to delay the triggering of an alert for some time when a vehicle is moving slowly, which could even end in an alert not triggered if the vehicle or the pedestrian moves away. In principle, we are not thinking of directly using the speed of vehicles to predict collisions, as this could affect the robustness of the algorithm against inaccuracies in location information and sudden changes in the speed of vehicles, generating false negatives. Another possibility is to consider the speed and the direction of movement of pedestrians, but given that they can change direction and speed very quickly, we do not believe this would be a good approach.

Furthermore, in our simulations, the maximum speed of pedestrians is 1.6 m/s, which is considered a realistic speed for pedestrians. This speed means that there is plenty of room to increase the beacon time period (we use 300 ms) for any reasonable speed of pedestrians (and even bicycles) while keeping accurate location information, even accounting for potential lost beacons. The advantage of increasing the beacon period is the saving in battery consumption in the pedestrian device and the reduction of network congestion [18]. For the same reason, we could use communications systems with more latency without that affecting the performance of the system.

## 6. Conclusion

In this paper, we have explored how to create a warning system to protect pedestrians in urban scenarios using a communication system that sends the location of pedestrians to vehicles. The main focus of our research has been on how to avoid the generation of too many unnecessary alerts in the system. The problem with unnecessary alerts is that they overload drivers, so they stop paying attention to alerts. This problem cannot be studied by defining a limited scenario to test the system (e.g., a junction with a small number of pedestrians). The load generated by the alerts must be tested in a large scale urban scenario, so any potential situation will eventually come up. It is on trips around a city where we can realistically study how pedestrians and vehicles interact, and how alerts are triggered. Therefore, and due to the practical limitations of performing these kinds of tests in the real world, we have proposed a realistic simulation framework based on a well-known accurate traffic simulator, SUMO.

We have designed four different alternative algorithms to define when to trigger alerts in the warning system. The algorithms use local information available in each vehicle, and the location of pedestrians obtained through the communication system. Alerts warn drivers on vehicles that they can become a danger to a pedestrian. The four algorithms use different strategies to determine when an alert is needed in a vehicle.

We have implemented the algorithms using real equipment: a communication and location module in vehicles, and a mobile phone with an application in pedestrians. Our experiments with these implementations have validated that it is possible to generate the alerts in the real world as defined by the algorithms, and that the errors introduced when using real equipment do not prevent the algorithms from working as expected. In particular, we have shown that the performance of the algorithms is robust against errors in determining the location in the mobile phone.

It should be noted that the system requires a constantly updated pedestrian location. This is a feasible approach in a hyperconnected world (e.g., IoT solutions or smartphones), but it is important to acknowledge that initial deployment of this kind of systems, which requires many users to be useful, is not straightforward. Other types of systems could use the same algorithms but based on different equipment. For example, an alternative is to equip crossings with devices that detect pedestrians with sensors or cameras, and send messages to vehicles. This would facilitate the initial deployment. Another deployment opportunity is to use dedicated devices for special interest groups such as children or the elderly. Privacy is another concern, so the communications should not use permanent identifiers that could allow long-term or repeated tracking of users.

Although the four proposed algorithms generate the alerts as expected in a small-scale setup with real-equipment, our simulation experiments have shown that their behavior in a large-scale deployment is much more difficult to predict. More specifically, increasing the complexity of the algorithms to try to better capture the real situations of danger for pedestrians does not always result in significant gains in the number of unnecessary alerts successfully filtered. In fact, algorithms that seem to achieve similar results when tested in a small-scale setup can have very different behavior in terms of number of unnecessary alerts that they generate. Therefore, to study the feasibility of a warning system to protect pedestrian is of paramount importance to evaluate the load on drivers generated by the alerts of the system, and study it when using the system in typical travels around an urban area.

Finally, one of the algorithms that we have proposed (algorithm 3) has proven to be a good first step to build a warning system to protect pedestrians in urban areas. Algorithm 3 is based on triggering alerts considering the distance between the vehicle and the pedestrian, but focusing only on situations that happen in front of the vehicle and in places where the pedestrians are more probably going to cross the road. The algorithm is realistically implementable with current technology, it does not lose any real alerts, and it does a good job filtering unnecessary alerts. Our algorithm only protects pedestrians at crossings or special interest places (e.g., intersections), but even so, the time drivers spent under alerts is not negligible. This fact gives the hint that realistic pedestrian protection systems for an urban environment cannot warn of any pedestrians in vehicles' surroundings. A better approach is to focus on places where pedestrians are more likely to cross the road. Our algorithm could be updated to warn about pedestrians leaving the sidewalk and entering the road when more precise location systems became available on pedestrians' devices.

Our ongoing research includes improvements to the algorithm, for example using the speed of vehicles to define a dynamic $th_{ad}$. We also plan to investigate the exploration of alternatives for the implementation of the mechanisms required by the algorithm (e.g., how to obtain the location of pedestrians), and the evaluation of practical strategies for deployment.

**Supplementary Materials:** The following are available online at www.mdpi.com/xxx/s1, Figure S1: title, Table S1: title, Video S1: title.

**Author Contributions:** Conceptualization, I.S., F.J., and M.C.; methodology, I.S., F.J., and M.C.; formal analysis, I.S.; experiments with real equipment, F.J., J.E.N., and J.A.; simulations, I.S.; analysis of results, I.S., F.J., and M.C.; writing—original draft preparation, I.S., F.J., and M.C.; writing—review and editing, I.S., F.J., and M.C.

**Funding:** The work in this article has been partially supported by the Spanish Ministerio de Economía y Competitividad (Ignacio Soto y Maria Calderon through the Texeo project, TEC2016-80339-R; and Felipe Jimenez, Jose E. Naranjo, and Jose J. Anaya through the CAV project, TRA2016-78886-C3-3-R).

**Conflicts of Interest:** The authors declare no conflict of interest. The funders had no role in the design of the study; in the collection, analyses, or interpretation of data; in the writing of the manuscript, or in the decision to publish the results.